# Influence of generalized focusing of few-cycle Gaussian pulses in attosecond pulse generation

**Ebrahim Karimi**[1,5,*], **Carlo Altucci**[1,2,†], **Valer Tosa**[3], **Raffaele Velotta**[1,2], **and Lorenzo Marrucci**[1,4]

[1] *Dipartimento di Fisica, Università di Napoli Federico II, Napoli, Italy*
[2] *Consorzio Nazionale Interuniversitario per le Scienze Fisiche della Materia – CNISM, Napoli, Italy*
[3] *National Institute for R&D Isotopic and Molecular Technologies, Cluj-Napoca, Romania*
[4] *CNR-SPIN, Complesso Universitario di Monte S. Angelo, Napoli, Italy*
5 *Currently with Department of Physics, University of Ottawa, Ottawa, Ontario, Canada*
\* Email: ekarimi@uottawa.ca
† Email: altucci@na.infn.it

**Abstract:** In contrast to the case of quasi-monochromatic waves, a focused optical pulse in the few-cycle limit may exhibit two independent curved wavefronts, associated with phase and group retardations, respectively. Focusing optical elements will generally affect these two wavefronts differently, thus leading to very different behavior of the pulse near focus. As limiting cases, we consider an ideal diffractive lens introducing only phase retardations and a perfect non-dispersive refractive lens (or a curved mirror) introducing equal phase and group retardations. We study the resulting diffraction effects on the pulse, finding both strong deformations of the pulse shape and shifts in the spectrum. We then show how important these effects can be in highly nonlinear optics, by studying their role in attosecond pulse generation. In particular, the focusing effects are found to affect substantially the generation of isolated attosecond pulses in gases from few-cycle fundamental optical fields.

## References and links

**1. Introduction**

The propagation and diffraction of ultrashort light pulses has been the subject of study for many years since the early pioneering papers [1, 2]. The importance of this issue is twofold: on one hand the last decade has been characterized by a knowledge boost in describing the intrinsic and unique optical properties of ultrashort light pulses; on the other hand such pulses have been widely employed in very many applications, from attosecond physics [3] to LIDAR applications in atmospheric physics [4, 5]. In particular, when dealing with extremely short pulses, i.e. in the few-cycle limit, the frequency-dependent nature of diffraction itself becomes a significant effect which can modify the pulse wavefront, shape and spectrum distributions, giving rise to effects which are strongly combined in space and time [6, 7]. In recent years diffraction effects in ultrashort pulses have also attracted a lot of interest when the spatial shape of the beam is somewhat exotic, such as X- or Bessel-Gauss beams [8], or for a possible superluminal behaviour arising in the on-axis group velocity because of Gouy phase shift with frequency [9]. Lately, it has been experimentally shown that few-cycle pulses can undergo dispersion-free and diffraction-limited focusing through all-reflective microscope objectives [10], and it has been predicted that diffraction spatial spreading of focused pulses does not occur in the limiting case of a 0-length, δ pulse [11].

Here we consider the effect of generalized focusing optical elements on few-cycle pulses and on their action in the highly nonlinear process of high-harmonic generation. In contrast to the case of quasi-monochromatic waves, a transmissive focusing element may introduce to first order two distinct effects on such pulses, respectively associated with phase and group time retardation. One can correspondingly define two independent curved "wavefronts": (i) the "phase wavefront", defined as the surface of constant field oscillation phase at a given time; (ii) the "group wavefront" (also called "pulse front"), defined as the surface of maximum amplitude (or intensity) of the pulse at a given time, as pictorially shown in Fig. 1. The group wavefront is, of course, related to the wavelength dispersion of the phase wavefront and they represent the two lowest-order effects of a more general dispersive focusing law. For a standard refractive lens, these two wavefronts will generally have different curvature radii. For example, a thin lens with a focal length $f = (n-1)(1/R_1 - 1/R_2)$, where $n$ is the refractive index and $R_1$ and $R_2$ the entrance and exit face curvature radii, will induce in an input plane-wave pulse an output phase wavefront curvature radius $R_p = f$ and a group wavefront curvature radius $R_g = f(N-1)/(n-1)$, where $N = n - \lambda(dn/d\lambda)$ is the group index. Of course this effect will not occur when using curved mirrors as focusing elements. The latter ones are usually chosen to minimize dispersion effects, and will always induce equal wavefront radii $R_p = R_g$. However, some applications have an advantage or even require using transmissive optical elements, and hence it is interesting to study these effects and to look for strategies for best controlling them. Among transmissive optical elements, besides standard refractive lenses, diffractive lenses (neglecting their wavelength dependence) will generally affect the phase wavefront only, leaving the group wavefront unaffected (or actually slightly tilted, if the diffraction angle is taken into account). Thus, a plane wave crossing a diffractive lens will have $R_p = f$ and $R_g = \infty$ at the output. This behavior is also shared by zero-order diffractive elements, such as Fresnel lenses and Pancharatnam-Berry phase optical elements [12]. Therefore, a strategy for the optimal focusing of ultrashort pulses in a transmission geometry could possibly benefit from using refractive-diffractive lens duplets, as first proposed by Ibragimov [13].

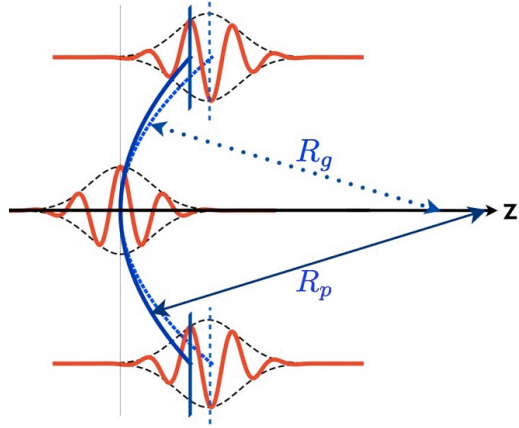

Fig. 1. Schematic description of the two distinct wavefronts characterizing, to first approximation, the effects of generalized focusing on a few-cycle optical pulse: the solid curve corresponds to the carrier-phase wavefront, following the location of the optical oscillation peak versus the transverse coordinates, while the dotted curve corresponds to the group (or pulse) wavefront, following the location of the pulse maximum. These two wavefront surfaces are characterized by the corresponding curvature radii $R_p$ and $R_g$, respectively.

To demonstrate the importance of these effects, we consider here the near-focus diffraction of few-cycle pulses that are focused by different optical elements. In particular, as main limiting cases, we consider an ideal diffractive lens introducing only phase retardations and an ideal non-dispersive lens (or a curved mirror) introducing equal phase and group retardations. We study the resulting diffraction effects on the pulse and found a very different behavior in the two cases. In particular, one significant finding is that the pulse spectrum is always modified in the focus, due to spatio-temporal interplay of diffraction effects in broadband pulses resulting in a spatial chirp of the final beam. The shorter the pulse length the stronger the spectral distortion. This may have serious consequences in several fields of application of few-cycle pulses, particularly in attosecond pulse generation [14] and in the general task of phase-matching in nonlinear optical processes [15-16]. In particular, as an outstanding example of the relevance of the above effects, we study here the influence of focusing of nearly single-cycle optical pulses in the generation of isolated attosecond pulses [17].

## 2. Few-cycle pulse propagation undergoing various types of focusing

We study the pulse propagation within a scalar paraxial approximation. The optical field in the input pupil plane $z = 0$, located just after the focusing element, is assumed to be given by the following expression:

$$E(\rho,\varphi,0;t) = A_0 e^{-\frac{\rho^2}{w_0^2} - i\omega_0(t+\varsigma_p\rho^2) - \left(\frac{t+\varsigma_g\rho^2}{\tau_0}\right)^2}, \qquad (1)$$

where $\rho$, $\phi$, $z$ are the cylindrical coordinates. This expression corresponds to a wavepacket having field amplitude $A_0$, carrier frequency $\omega_0$, Gaussian pulse shape with duration $\tau_0$, and Gaussian beam transverse profile with waist $w_0$. Moreover we have introduced a parabolic-shaped phase wavefront, controlled by the coefficient $\varsigma_p$, and a group wavefront, controlled by the coefficient $\varsigma_g$. These coefficients are related to the above defined wavefront curvature radii by $\varsigma_p = 1/(2cR_p)$ and $\varsigma_g = 1/(2cR_g)$, where $c$ is the light speed in vacuum. In the frequency domain, the input field is then given by the following expression:

$$\tilde{E}(\rho,\varphi,0;\omega) = \frac{1}{\sqrt{2\pi}} \int E(\rho,\varphi,0;t) \cdot e^{i\omega t} dt = \frac{A_0 \tau_0}{2\sqrt{2}} e^{-\frac{\rho^2}{w_0^2} - i\omega_0\varsigma_\omega\rho^2 - \frac{\tau_0^2}{4}(\omega-\omega_0)^2}, \qquad (2)$$

where we have set $\varsigma_\omega = \varsigma_p + (\omega-\omega_0)\varsigma_g/\omega_0$. Each frequency component behaves as a monochromatic Gaussian beam and can be therefore propagated to arbitrary $z$ by using the standard Gaussian beam propagation laws. This leads to the following simple expression for the propagated field in the frequency domain:

$$\tilde{E}(\rho,\varphi,z;\omega) = \frac{A_0\tau_0}{2\sqrt{2}}\alpha(z,\omega)e^{i\frac{\omega}{c}z - \alpha(z,\omega)\left(1+i\omega_0\varsigma_\omega w_0^2\right)\frac{\rho^2}{w_0^2} - i\omega_0\varsigma_\omega\rho^2 - \frac{\tau_0^2}{4}(\omega-\omega_0)^2}, \qquad (3)$$

where $\alpha(z,\omega)$ is related with the Gaussian beam $q(z)$ parameter and is given by:

$$\alpha(z,\omega) = \frac{q(0)}{q(z)} = \frac{\omega w_0^2}{\omega w_0^2 + 2cz(i - \omega_0\varsigma_\omega w_0^2)}. \qquad (4)$$

This expression must be finally Fourier-transformed back in the time domain to obtain the propagated field in space and time. We perform this last step by numerical integration.

Two specific limiting cases are then analyzed: (i) focusing by a flat diffraction optical element of focal length $f$, corresponding to taking $\varsigma_p = 1/(2cf)$ and $\varsigma_g = 0$; (ii) focusing by an ideal non-dispersive refractive lens or curved mirror of focal length $f$, for which $\varsigma_p = \varsigma_g = 1/(2cf)$. For definiteness, we consider a few-cycle pulse having a central frequency $\omega_0 = 3.8\times10^{15}$ rad/s, beam waist of $w_0 = 1$ mm, pulse duration of $2\tau_0 = 4$ fs, and a focal length $f = 50$ cm. The beam radius in the focal plane is $a_0 = w_0\left(1 + f^2/z_R^2\right)^{-1/2}$ where $z_R = \omega_0 w_0^2/(2c)$. Fig. 2 shows the field amplitude in the time-domain (left column in each panel, blue line) and frequency-domain (right column in each panel, red line) in the two mentioned limiting cases. It can be seen that the temporal pulse shape is deformed, with a distortion that increases with the distance from the beam axis. The case of focusing by an ideal non-dispersive lens (or mirror) leads to effects similar to those already studied in the case of free propagation [1, 2]. The distortions are instead much stronger in the case of focusing by an ideal diffractive lens. In all cases, however, significant modifications are induced to the pulse spectrum. In the case (i), a small blue-shift occurs exactly on the beam axis, i.e. the central pulse frequency increases with respect to the input field. A spatial chirp is also induced, since the spectrum exhibits a significant red-shift at the beam periphery. In the case (ii), the pulse spectrum at the beam periphery even splits in two bands which become more and more separated with increasing radius. This effect is similar to the spectrum modifications of polychromatic continuous waves at focus reported in [18-20], where the spectrum changes around a phase singular point.

The spectral blue and red shifts suffered by a pulse on focusing are a simple consequence of the frequency-dependent diffraction (even for perfectly non-dispersive lenses), as blue spectral components are focused more tightly than red ones and hence will have a greater weight in the central region of the beam, while the periphery will collect a major contribution of red components. Fig. 3 shows the shift of the peak frequency of the wave spectrum as a function of radius for the specific choice of parameters mentioned above.

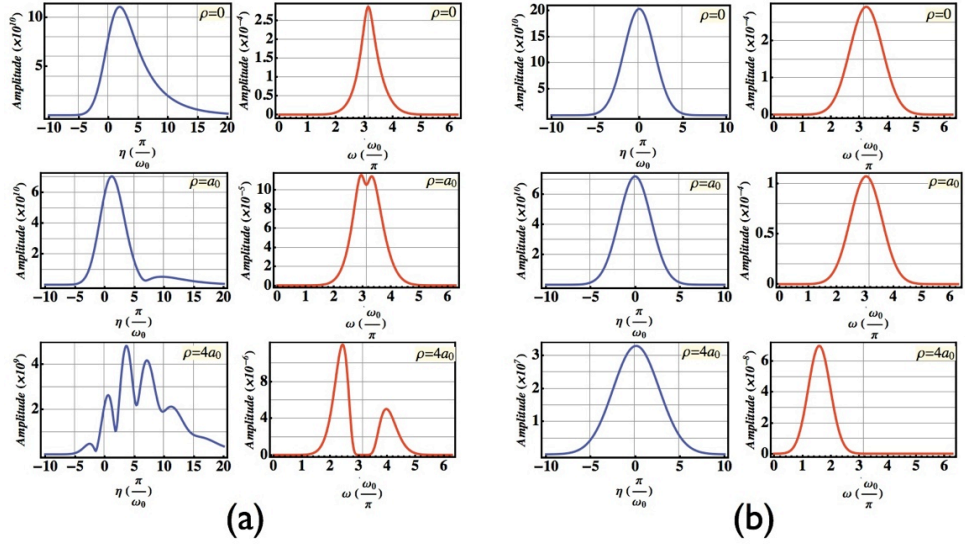

Fig. 2. Left (blue) and right (red) side of each column show the pulse and spectrum shape of optical field amplitude (square root of the energy density) for (a) focusing by an ideal diffractive lens and (b) focusing by an ideal non-dispersive lens (or a curved mirror) on axis ($\rho = 0$) and at various distances $\rho$ from the beam axis, relative to the beam waist $a_0$ in the focal plane. See text for the choice of parameter values. $\eta = t - z/c$ is the retarded time at the given $z$..

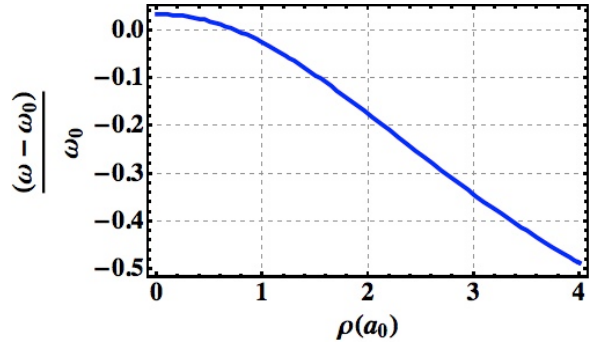

Fig. 3. Relative shift of the spectrum peak for a Gaussian pulse beam at the focus of an ideal non-dispersive lens (or focusing mirror) as a function of radius in units of the focal-plane beam waist $a_0$.

## 3. Applications to single attosecond pulse generation

The results presented in Sec. 2 demonstrate that, when focusing few-cycle pulses, the spatial and spectral/temporal configurations increasingly deviate from the "standard" configurations of many cycles pulses and continuous-wave (cw) fields. For extremely short pulses, such as the near single cycle pulses [17] or the recently generated synthesized light transients [21], this deviation will clearly affect our description and understanding of the interaction of these short waveforms with matter. We investigate here a specific case, namely the generation of high-order harmonics following the interaction of gas phase targets with few cycle pulses. Describing the macroscopic response of the medium in this interaction is relevant because in most cases single dipole response is not sufficient in explaining the experimental data. Indeed, the peculiarities observed in the configuration of the beam across the focal region can play an essential role in shaping the spatial configuration of the propagated laser field in the ionized medium and ultimately in building-up the harmonic field in the medium.

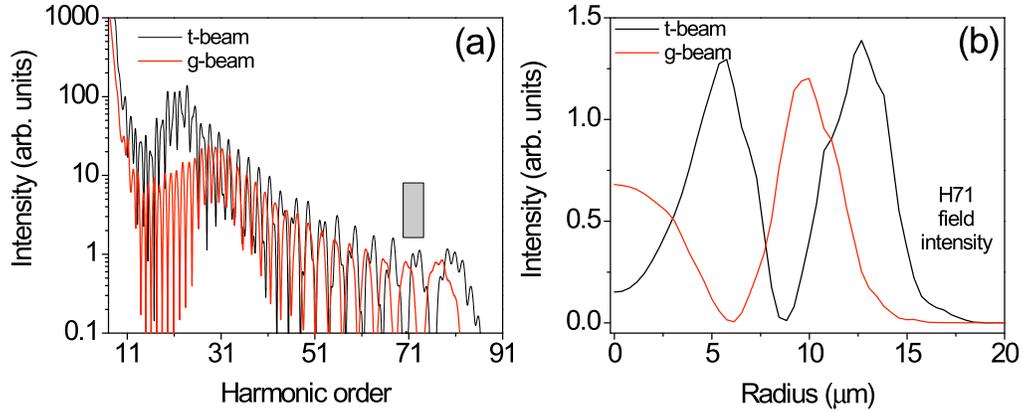

The problem of laser field and harmonic field propagation in an ionized medium was solved numerically in a three dimensional (3D) non-adiabatic model [22], which is particularly apt to this case, because it solves the wave equation for the full electric field describing the laser and harmonic pulses at optical cycle level. The single dipole response was calculated within the Strong Field Approximation [23] using as driving field the propagated laser field. The expression for the pulse field as given by Eq. 3 was used to calculate the field at the entrance of the interaction region, which were subsequently utilized as initial values to integrate the wave equation and eventually to obtain the macroscopic harmonic field as described above.

Fig. 4. (a) Spectra of single dipole response for the two 1.8 fs pulses, on-axis. The shaded grey area indicates the spectral region used to plot the radial dependence of dipole intensity shown in (b).

We analyzed the generation of harmonics in a 2 mm long medium filled with 40 Torr of He, illuminated by a sub-cycle pulse of 1.8 fs duration (FWHM taken for the envelope) focused by a mirror of 50 cm focal length. The intensity in the focus reaches $2.2 \times 10^{15}$ W/cm$^2$ and induces an ionization fraction of ~18% which is covering a typical case of high intensity high order harmonic generation experiment. In a first calculation, we used the pulse field configuration as given by Eq. 3 to provide the initial values for pulse propagation in the ionized medium. For comparison purposes we performed a second calculation by assuming a 1.8 fs pulse, Gaussian in both time and radial direction and having the same central wavelength of 800 nm. The latter corresponds to what one would expect to have when neglecting the diffraction effects of the focusing element, as legitimate in the many-cycle limit. The two cases will be hereafter called t-beam (to suggest a somehow "true"-beam) and g-beam, respectively.

The differences between the two cases, as seen in Fig. 4, start even at single dipole response, before considering propagation in the medium: one can find, for certain frequencies, even one order of magnitude difference in intensity. The radial dependence of the dipole response for a spectral region around the 71$^{st}$ harmonic (H71) is also shown in Fig. 4 (b) and is clearly different for t- and g-beams. It is therefore expected that the final harmonic field will be dissimilar for the two cases. Indeed, the total power spectrum (radially integrated harmonic intensity) generated at the exit from the interaction region, (see Fig. 5) is markedly different both in intensity and in cutoff. While the cutoff of the single dipole is roughly the same in Fig. 4, the macroscopic power spectra exhibit different values for the cutoff. This is an indication that the phase-matching conditions are uneven for the two cases, which is reflected also in the intensity values.

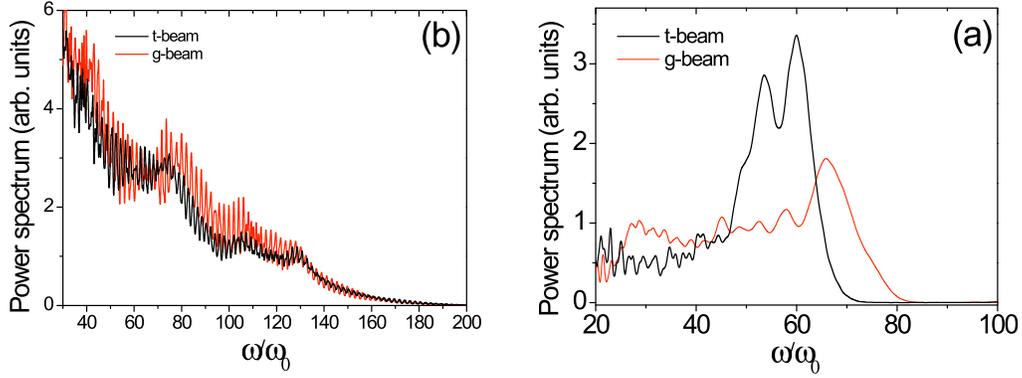

Fig. 5. Power spectra of harmonics generated by the 1.8 fs pulses (a) and 4.5 fs pulses (b).

The same calculation was also performed for few-cycle pulses of different durations. As expected, we note smaller differences between t- and g-beam cases for larger pulse durations, starting from driving field values and ending with final harmonic field. The 4.5 fs case is shown in Fig. 5 (b) for the power spectrum. One can still note differences, but the trend and the cutoff of the spectra are roughly the same. One can say that the differences become entirely irrelevant for sufficiently long pulses.

It is certainly interesting to see how the temporal structure is affected by considering the true configuration for the focused field or just the Gaussian configuration, as shown in Fig. 6. Here, the 1.8 fs and 4.5 fs harmonic spectra were filtered out in their cutoff part by the same bandwidth (30 and respectively 50 harmonic orders) and the inverse Fourier transform was then applied. Noticeable differences are found in the attosecond pulse durations: for 1.8 fs case we get a single attosecond pulse lasting 130 and 290 as for the s-beam and g-beam, respectively, while for the 4.5 fs case we obtain 160 and 120 as.

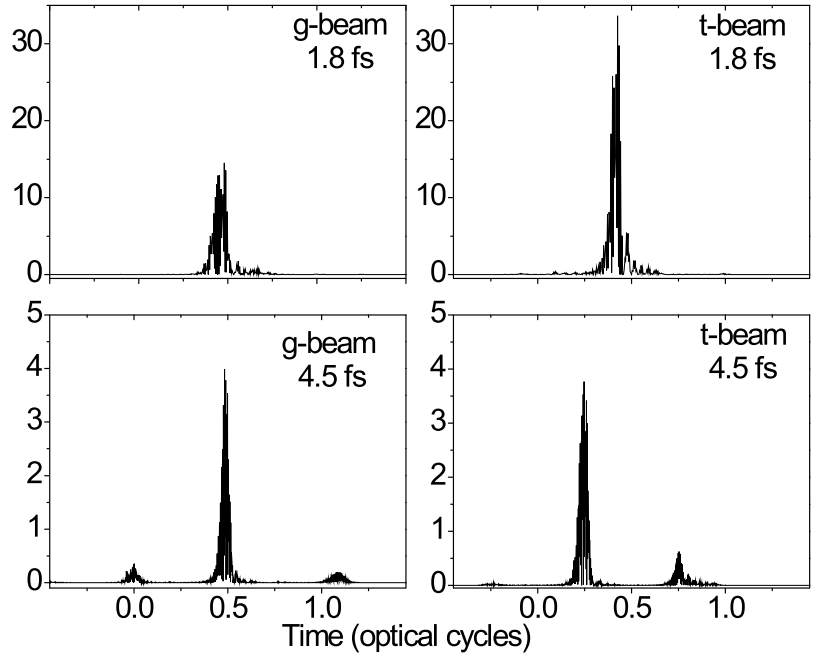

Fig. 6. Single attosecond pulses obtained by filtering the cutoff of the spectra in Fig. 5. Vertical axis in arbitrary but same units in all panels.

## 4. Conclusion

In conclusion, we have discussed a generalization of the effects of focusing optical elements which is relevant in the few-cycle pulse regime. We have shown that the resulting behavior of the optical pulse near focus is strongly affected by the specific properties of the focusing element. In particular, significant spectral shifts depending on the transverse coordinates occur in all cases, even in the simplest non-dispersive ideal focusing. These effects are usually not accounted for and may have important fallouts in few-cycle pulse applications such as attosecond science and highly nonlinear optics.

As a significant example of the effects here described, we have simulated a typical single attosecond pulse generation experiment, based on high harmonic generation in He jets by 4.5 and 1.8 fs focused laser pulses centered at 800 nm carrier wavelength. We have compared the outcome when using the few-cycle pulse obtained by the true complete propagation, for which spatial and time evolution of the pulse are interconnected, with the result obtained with the typical approximation of a few-cycle field, for which the spatial propagation is decoupled from the pulse temporal evolution. The comparison is extremely interesting, showing the essential role played by the "true" structure of the wave in space and time when the shorter fundamental pulse is used. Our investigation shows that, for the sake of a correct understanding of high-harmonic and attosecond pulse generation, an appropriate treatment of focusing and propagation of near-single-cycle and sub-cycle pulses cannot be done without taking into account the real wavefronts, which are being formed and propagate in the diffraction process. In particular, for modeling high-harmonics generation, considering the real wavefront brings a better description of the physical process, basically because the shape of the field in space will influence the phase-matching process in both axial and radial direction. In time domain, the initial wavefront curvature has an influence on the formation of attosecond pulses, either in trains or single pulses. Indeed, the ionization strongly varies in radial direction and induces additional wavefront deformations during propagation so that the bursts have a continuous shift in emission time [21] along radial direction. The structure of the initial wavefront of the fundamental field is found to be essential since it can help or deteriorate the formation of short attosecond pulses. The simulations also show that the spectra of the generated XUV radiation are different in the two cases, indicating that both the spectral extension and the efficiency of the conversion into XUV radiation can significantly differ.

## Acknowledgements

E.K. and L.M. acknowledge the financial support of the Future Emerging Technologies (FET)-Open Program within the 7[th] Framework Programme of the European Commission under the Grant No. 255914, Phorbitech. C.A., and R.V. acknowledge the financial support of the European Community under the Grant No. 221952, ATLAS, within the 7[th] Framework Programme of the European Commission. V.T. acknowledges partial support partial support from UEFISCDI through the contract PN-II-ID-PCE-2012-4-0342.